\documentclass[]{scrartcl}
\usepackage{cuted}
\usepackage{breqn}
\usepackage{bm}
\usepackage{makecell}
\usepackage{url}
\usepackage[english]{babel}
\usepackage{amsfonts}
\usepackage{latexsym}
\usepackage{graphicx}
\usepackage{graphics}
\usepackage{float}
\usepackage{hyperref}
\usepackage{amssymb}
\usepackage{authblk}
\usepackage{multirow}
\usepackage{makeidx}
\usepackage{dblfloatfix}
\usepackage[section]{placeins}
\usepackage{flushend}
\usepackage{microtype}
\usepackage{breqn}
\PassOptionsToPackage{hyphens}{url}

\begin{document}
\title{A new laser-ranged satellite
for General Relativity and Space Geodesy
}
\subtitle{IV. Thermal drag and the LARES 2 space experiment }

\author[1,2]{Ignazio Ciufolini\thanks{ignazio.ciufolini@unisalento.it}}
\author[3]{ Richard Matzner}
\author[3]{Justin Feng}
\author[4]{David P. Rubincam}
\author[5]{ Erricos C. Pavlis}
\author[6]{ Giampiero Sindoni}
\author[6]{ Antonio Paolozzi}
\author[2]{Claudio Paris}

\affil[1]{\footnotesize  Dip. Ingegneria dell'Innovazione, Universit\`a del Salento, Lecce, Italy}
\affil[2]{Museo della fisica e Centro studi e ricerche Enrico Fermi, Rome, Italy}
\affil[3]{Theory Group, University of Texas at Austin, USA}
\affil[4]{NASA Goddard Space Flight Center, Greenbelt, Maryland, USA}
\affil[5]{Goddard Earth Science and Technology Center (GEST), University of Maryland, Baltimore County, USA }
\affil[6]{Scuola di Ingegneria Aerospaziale , Sapienza Universit\`a di Roma, Italy }

\renewcommand\Authands{ and }

\date{}
\maketitle

\abstract{
In three previous papers we presented the LARES 2 space experiment aimed at a very accurate test of frame-dragging and at other tests of fundamental physics and measurements of space geodesy and geodynamics. We presented the error sources in the LARES 2 experiment, its error budget, Monte Carlo simulations and covariance analyses confirming an accuracy of a few parts per thousand in the test of frame-dragging, and we treated the error due to the uncertainty in the de Sitter effect, a relativistic orbital perturbation. Here we discuss the impact in the error budget of the LARES 2 frame-dragging experiment of the orbital perturbation due to thermal drag or thermal thrust. We show that the thermal drag induces an uncertainty of about one part per thousand in the LARES 2 frame-dragging test, consistent with the error estimates in our previous papers.

} 
\section{Introduction: thermal drag and LARES 2}
\label{intro}
We recently described a new satellite experiment to measure the General Relativistic phenomenon know as {\itshape frame dragging}, or the Lense-Thirring effect. The idea of the experiment is to accurately measure the location of the ascending node\footnote{For a geocentric orbit such as LARES, the longitude of the ascending node (also known as the right ascension of the ascending node) is the angle, measured eastward, from the vernal equinox to the ascending node. The ascending node is defined to be the intersection of LARES's orbit with the Earth's equatorial plane, where LARES is heading north.} of the satellite(s).

We analyze the uncertainty in our proposed measurement of frame-dragging with LAGEOS and LARES 2 arising from anisotropic thermal radiation from these satellites. When considering thermal effects driven by solar radiation, this phenomenon is known as the Yarkovsky, or Yarkovsky-Schach, or solar Yarkovsky effect. But additionally the LAGEOS (and proposed LARES 2) fused-silica retroreflectors are strong absorbers of Earth infrared radiation. The forces generated by the retroreflectors' anisotropic reemission is a phenomenon known as the Yarkovsky-Rubincam, or Earth-Yarkovsky effect \cite{bib1,bib2,bib3}.

The Yarkovsky effect (solar or Earth driven) arises as follows. The electromagnetic radiation from the Sun and the radiation from Earth each instantaneously heat one hemisphere of LAGEOS. Because of the finite heat conductivity of the body, there is an anisotropic distribution of temperature on the satellite. Thus, according to the Stefan-Boltzmann law, there is an anisotropic flux of energy ($\sim T^{4}$) and momentum from the surface of the satellite giving rise to its acceleration. If the satellite is spinning fast enough, the anisotropy in the satellite temperature distribution is mainly latitudinal, but that is not currently the case for LAGEOS, which currently has very slow rotation (compared to its orbital period \cite{bib4}) with respect to inertial space. Also, we assume that the LARES 2 satellite will be almost non-rotating at the moment of its release.
Thus we study the Yarkovsky (solar) and Yarkovsky-Rubincam effect (Earth) in the hypothesis of satellite that is not spinning with respect to inertial space.

\section{The Yarkovsky-Rubincam (Earth Yarkovsky) effect}

One of us \cite{bib1} discovered that the infrared radiation from the Earth plus the thermal inertia of the LAGEOS retro-reflectors can result in a force on the satellite. Infrared radiation from Earth is absorbed by the LAGEOS retro-reflectors, then reradiated anisotropically.  This thermal reradiation may cause an acceleration of LAGEOS of the order of $10^{-12} m/s^{2}$. It is present even when the satellite is not rotating; see Figure 1, which shows the geometry of the Yarkovsky-Rubincam Earth thermal drag for no rotation of a satellite of LAGEOS type. This figure shows how the infrared radiation from Earth generates a thermal anisotropy producing an along-track acceleration of the satellite but no out-of-plane acceleration, and thus there is no way this force to torque the orbit, and no way to change the location of the node.

\begin{figure}[b]
\centering
 \includegraphics[width=0.640\textwidth]{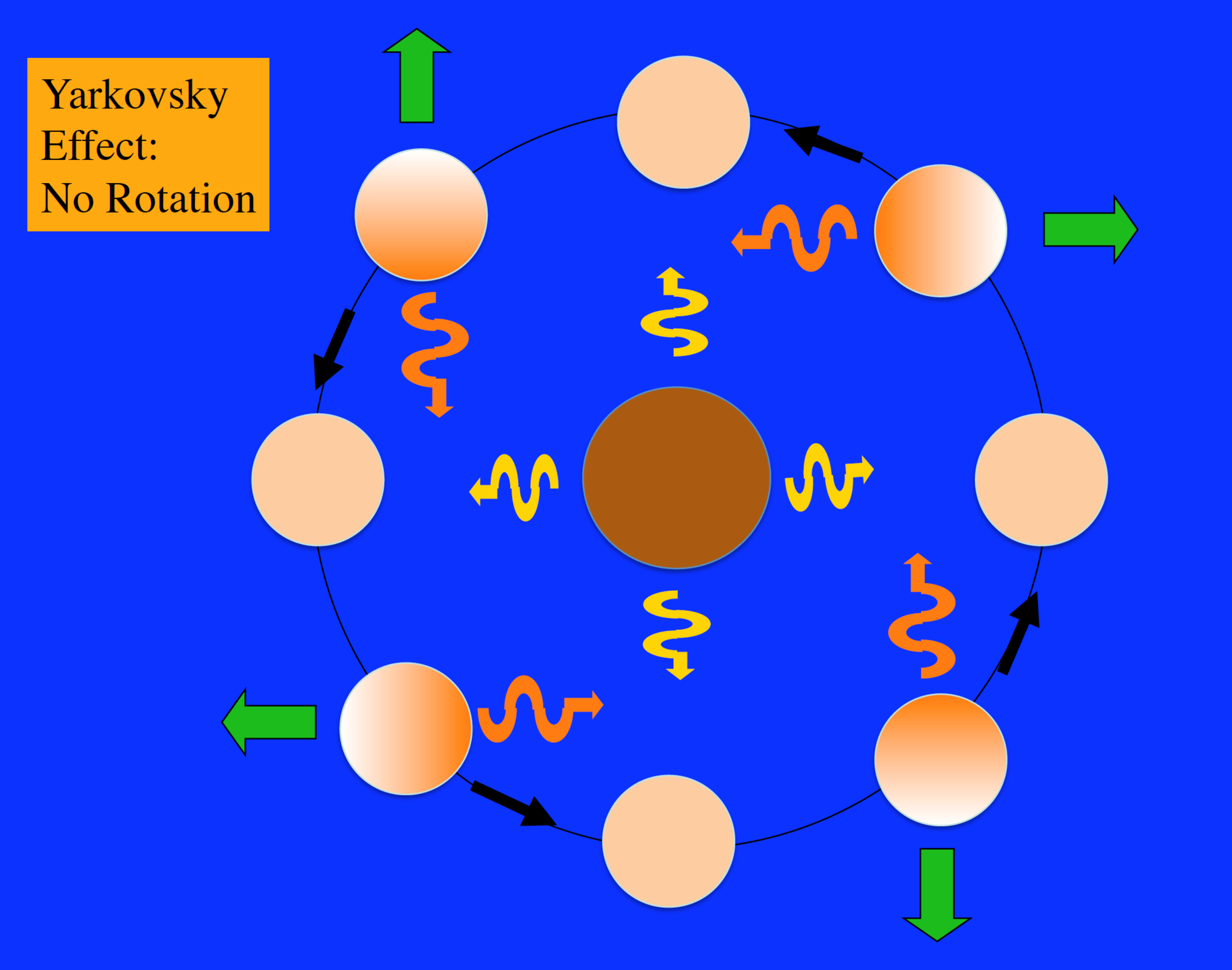}
\caption{The Yarkovsky-Rubincam effect (Earth Yarkovsky) for a non-rotating satellite. Earth is the brown sphere in the center.  LAGEOS, or LARES, orbits the Earth in the counter-clockwise sense (black arrows). Starting at the top:  Photons from the Earth (yellow) heat up the bottom of the satellite.  Because of thermal inertia, the bottom does not get hottest until after it has moved around the orbit.  As a result, the satellite radiates  most of the energy (red) downward at the next position as shown. This results in a kick on the satellite in the up direction (green arrow).  The kick has an along-track component in the direction opposite to the motion, acting as drag on the satellite. And so on all the way around the orbit, as shown.  Hence there is a secular drag, bringing the satellite closer to Earth. However, there is no out-of-plane component. (Figure courtesy of David Rubincam). }
\label{fig:1}       
\end{figure}

\section{The solar Yarkovsky effect}

The solar Yarkovsky effect is more complicated.  In the simpler (Yarkovsky) case the orbit does not intersect the Earth's shadow. In the case of a satellite non-rotating with respect to inertial space, the hemisphere looking at the Sun gets hotter and thus there is an additional anisotropic flux of thermal radiation and a recoil acceleration in the direction Sun-Earth. This force is directed away from the Sun, and simply adds to the solar radiation pressure. There {\itshape is} an out-of-plane component, which arises whenever the satellite orbit angular momentum is not perpendicular to the Sun-Earth direction. If orbital ellipticity is nonzero this out-of-plane force can torque the orbit and shift the location of the node, thus potentially introducing an error in measurement of the frame-dragging effect.. Because of its rapid classical nodal motion eastward ($\sim 0.343^\circ/day$), LAGEOS completes a complete cycle of Sun-orientation in $\sim 540$ days, so the direction and magnitude of the torque oscillates with this period. The value and sign of $d\Omega/dt$ also depend on the direction to the periapse. According to the Gauss equation [REFERENCE], the rate of change with time, $t$, of the nodal longitude $\Omega$ is:

\begin{dmath}\label{eq1}
  \frac{d\Omega}{dt}=\frac{rW\sin(\omega+f)}{h \sin i},
\end{dmath}
where $r$ is the satellite's radial distance, $h$ its orbital angular momentum per unit mass, $i$ its orbital inclination, $\omega$ its argument of pericentre, $f$ its true anomaly and $W$ the out-of-plane component of a force per unit mass acting on the satellite (the component of a perturbing force which acts orthogonally to the satellite orbital plane).  $ d\Omega/dt$ vanishes if the eccentricity does. (For zero eccentricity -- and only in this case -- $ f $ increases linearly in time, and  $d\Omega/dt $ averages to zero over the circular orbit.)

The complication arises when the orbit {\itshape does} intersect the Earth's shadow (the Yarkovsky-Schach effect).  See Figure 2. These {\itshape eclipse seasons} begin approximately every 270 days. [REFERENCE Slabinski 1977] Due to the satellite's thermal inertia, the satellite takes time to cool down in the shadow, and then takes time to warm back up after it exits the shadow.  The lag gives effects on the orbit.  One effect is to increase the semimajor axis (see Figure 2).  In general, one will get radial and along-track forces with Yarkovsky-Schach because they depend on the Sun-satellite orbit geometry. Out-of-plane forces can also arise as the satellite enters eclipse ``season", and in the opposite direction as it exits eclipse season. Note that nonzero $d\Omega/dt$ can result in this case even if the ellipticity is zero.

\begin{figure}
\centering
 \includegraphics[width=0.640\textwidth]{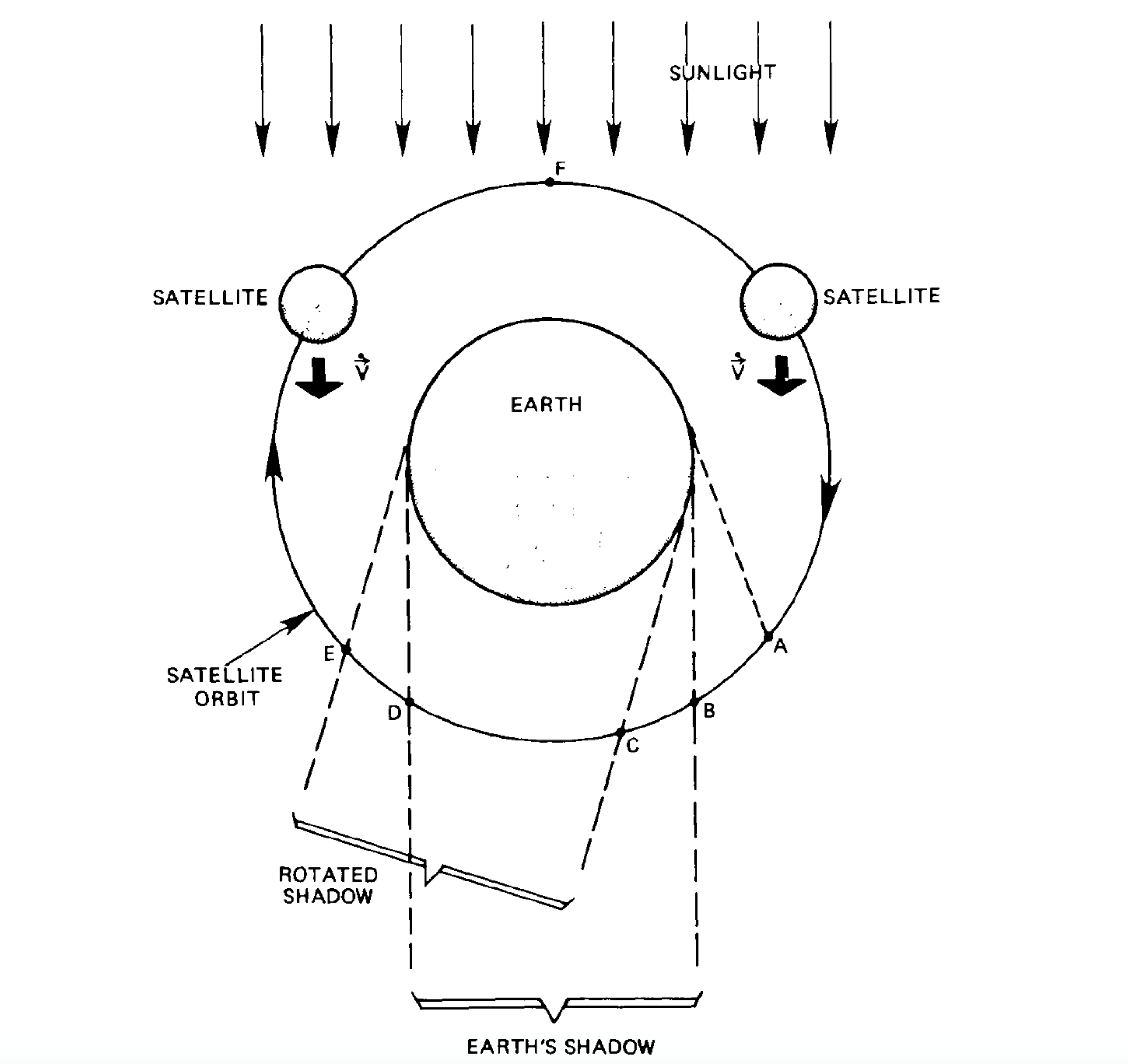}
\caption{The Yarkovsky-Schach effect. A satellite is shown at two points along its orbit. The acceleration $\dot{\vec{v}}$ is in the direction of the sunlight when the satellite is not in shadow. If the satellite cools off and warms up instantly when moving through the shadow, then the acceleration is zero in the shadow and the acceleration over arc FED cancels that over arc FAB by symmetry. However, if the satellite has some thermal inertia, then it takes some time to cool off and heat up when moving through the shadow. We may think of this effect as a thermally inertialess satellite passing through a rotated shadow inside of which $\dot{\vec{v}}$ is zero. In this case the acceleration over arc FE cancels that over arc FA, but no cancellation occurs for arc ABC. This leads to a net acceleration when averaging over one revolution which tends to increase the semimajor axis. The uncancelled force can in some cases also torque the orbit, even for circular orbits. Figure adapted from from \cite{bib5}}
\label{fig:2}       
\end{figure}

We begin with the non-eclipse situation. When the orbit of the satellite is not in the shadow of the Earth, the Yarkovsky acceleration corresponds to a change in the direct solar radiation pressure and is thus equivalent to a change in the $C_{r}$ of the satellite.
$C_{r}$ is a factor with value between 0 and 2, depending on the reflection properties of the satellite and on its geometry relative to the radiation source. $C_{r}$ determines the force on a satellite due to the radiation pressure and thus the solar-induced acceleration:  $a_{S} = C_{r} (A_{L}/m_{L}) \Phi/c$.
 $\Phi/c \cong  4.65 \times 10^{-6}N/m^{2}$ is the solar radiation pressure at the Earth
when the geocentric distance is equal to its mean distance, and $c$ is the speed of light. Specifically for LAGEOS, $C_{r}\cong 1.13$ and $A_{L}/m_{L}$ is the LAGEOS (small) cross-sectional-to-mass ratio equal to about $0.0007 m^{2}/kg$.   This leads to a LAGEOS acceleration due to the direct solar radiation pressure:
  $a_{S} \approx 3.7 \times 10^{-9}  m/s^{2}$.
For LAGEOS the solar Yarkovsky effect (the no-eclipse case) $a_{SY}$ is   $a_{SY} = 4 \epsilon \pi  r_{L}^{2} \sigma T_{0}^{3} \Delta T|_{eff}/(c \cdot m_{L}$), where $\epsilon \cong 0.4$ is the emissivity of LAGEOS, $r_{L} = 0.3 m$ and $m_{L} = 411 kg $ are the radius and mass of LAGEOS, and $\sigma = 5.67 \times 10^{-8} W m^{-2} K^{-4}$ is Stefan's constant. The temperature $T_{0} = 280^\circ C$ is the estimated average equilibrium temperature of LAGEOS,  and $\Delta T|_{eff} = 5^\circ$ is the estimated effective temperature difference between the two LAGEOS hemispheres.
The estimates lead to $a_{SY}= 2.28 \cdot 10^{-11} m/s^{2}$.

In order to estimate the effect of the solar Yarkovsky effect, in the case of no eclipses and no rotation of LAGEOS with
respect to inertial space, we can assume that the thermal drag would correspond to a change of the $C_{r}$ coefficient of about $(2.28 \cdot 10^{-11} m/s^{2}) / (3.7 \times 10^{-9}  m/s^{2}) = 0.6\% $. As explained above, in the case of no eclipses of the satellite by the Earth, the solar Yarkovsky effect will be in the same direction as the direct solar radiation and will simply add to the radiation pressure.

We performed, using GEODYN, data analyses of the real SLR data of LAGEOS in two cases: (a) assuming the standard value of the $C_{r}$  of LAGEOS of 1.13 and (b) assuming a variation of its $C_{r}$  of about $ 0.6\% $, that is a LAGEOS $C_{r}^{*} \cong  1.136$. However we subsequently allowed $C_{r}$  to be estimated over each run, since that is the procedure in our standard data analysis. In the end we calculated the variation, between these two cases, of the measured frame-dragging effect. The result is shown in Figure 4.

Over 2.65 years, we found an effect on the node of LAGEOS of about $5 \times 10^{-8}$ degrees, that is: $5 \times 10^{-8}\times 3600 \times 1000 mas / 2.65 yr = 0.068 mas/yr $, that is $ (0.068 mas/yr) / (31 mas/yr) = 2.2 \times10^{-3}$ fractional accuracy in comparison with $31  mas/yr$ of the frame-dragging shift on LAGEOS.

We emphasize that this result is proportional to the LAGEOS eccentricity of $\sim 4.5\times 10^{-3}$.

\begin{figure}
\centering
 \includegraphics[width=0.640\textwidth]{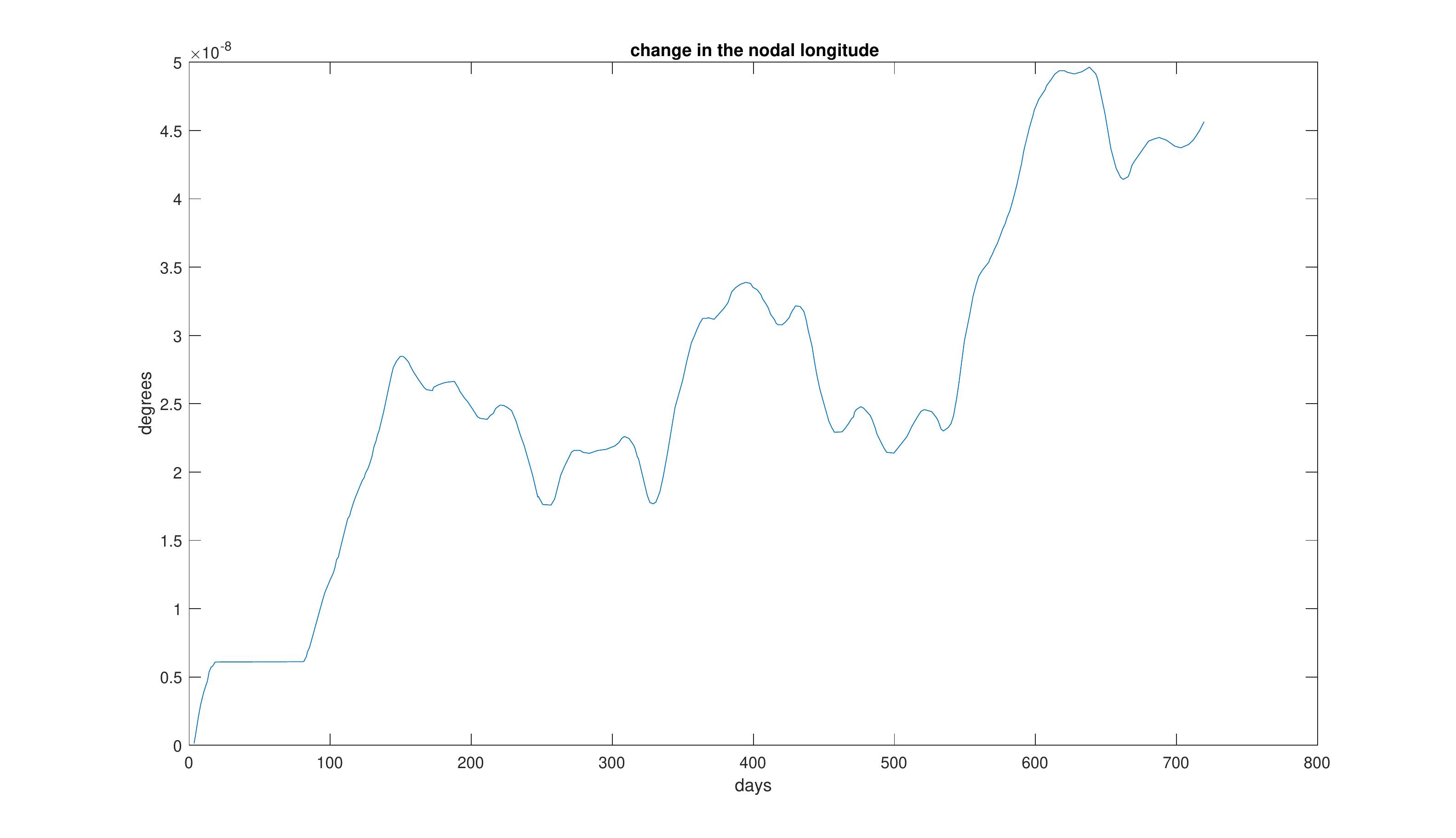}
\caption{Change in the nodal longitude of LAGEOS over about 1.97 years corresponding to a change to its $C_r$ of $0.6\%$ (see text for explanation). The total scale in the coordinate is $5\cdot 10^{-8}$ degrees. }
\label{fig:3}       
\end{figure}

Now consider the LARES 2 satellite. LARES 2 will have a cross-sectional to mass ratio about 1.5 times smaller than LAGEOS. In addition, LARES 2 will be machined from a single block of metal so temperature gradients across it will be much smaller than those on LAGEOS: $\Delta T_{LARES}< 1K$. (LAGEOS is composed of three components bolted together: two aluminum shells and a brass core. It is notorious for apparently poor thermal contact between its components, whence the large $\Delta T_{LAGEOS} \sim 5K$.) Design criteria call for LARES 2 to have an eccentricity comparable to that of LARES, roughly half of LAGEOS', but even if its eccentricity is the same magnitude as LAGEOS’, LARES' contribution to the error budget due to this effect is at the $0.3\times10^{-3}$ level, negligible compared to the LAGEOS error.

The logic of the frame dragging experiment is to add half the total LAGEOS nodal precession to half the total LARES 2 nodal precession. According to the uncertainty just calculated the associated ``half of LAGEOS'' contribution to the frame dragging uncertainty is $\frac{1}{2} (2.2 \times 10^{-3})$ ; the ``half of LARES 2'' contribution is $\frac{1}{2} (0.3 \times 10^{-3})$. We treat these as uncorrelated, so we add them in quadrature, yielding
\begin{dmath}
\bigg(1.21\times 10^{-6} + 0.023\times 10^{-6}\bigg)^{\frac{1} {2}} = 1.11\times 10^{-3}.
\end{dmath}
Thus when there are no eclipses, radiation pressure contributes errors at the $0.1\%$ level, consistent with our previous estimtes.

\begin{figure}
\centering
 \includegraphics[width=0.640\textwidth]{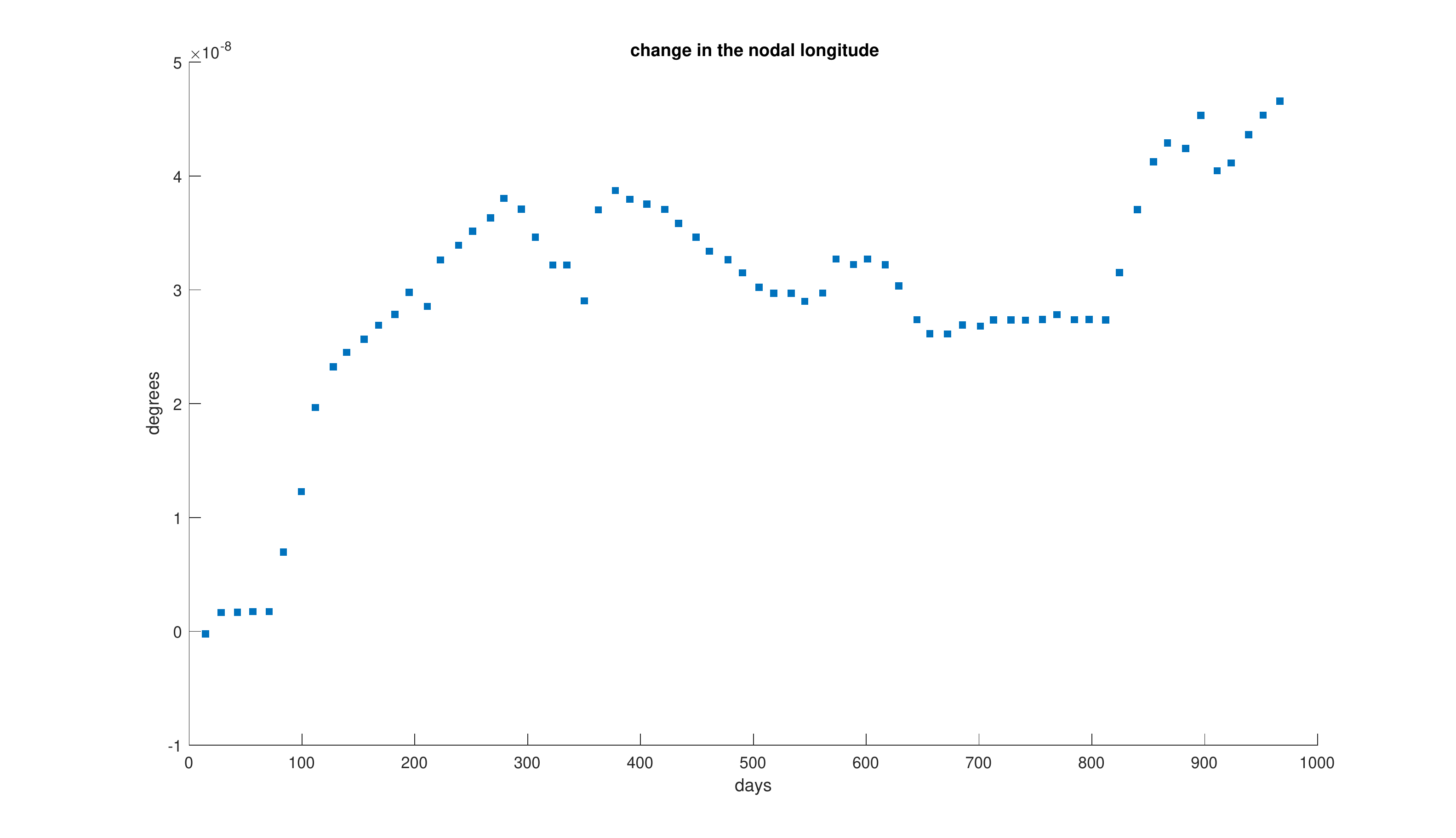}
\caption{Difference in the residuals of the nodal longitude of LAGEOS between the two cases of $C_{r}$ = 1.13 and $C_{r}$' = 1.136, over about 2.65 years, estimating the variations of the $C_{r}$ with GEODYN. Each residual was calculated over a 14-days arc.}
\label{fig:4}       
\end{figure}

\begin{figure}
\centering
 \includegraphics[width=0.640\textwidth]{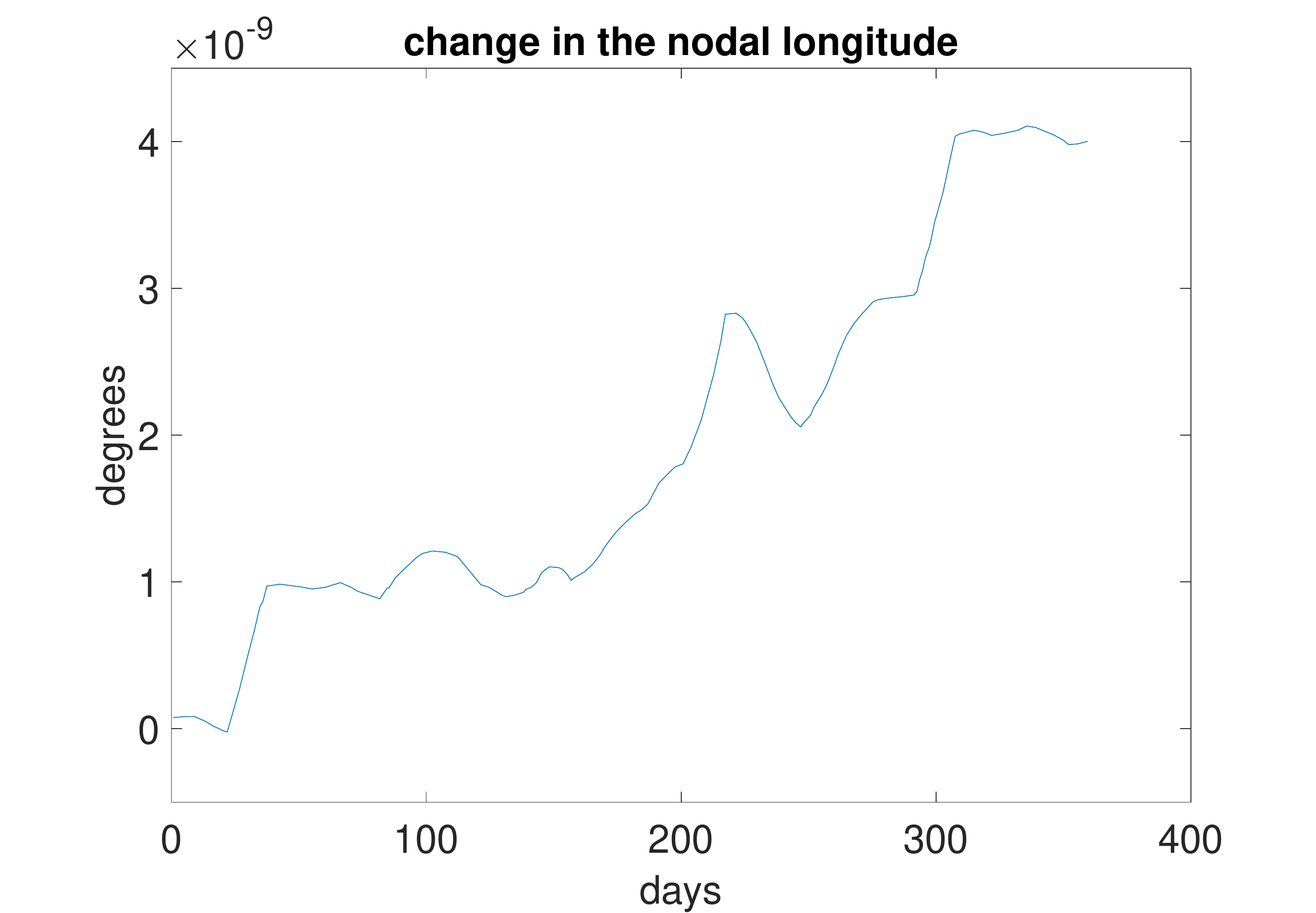}
\caption{Difference, over about 1 year, in the simulated residuals of the nodal longitude of LAGEOS between the two cases of no thermal thrust and a thermal thrust always directed towards the Sun.}
\label{fig:5}       
\end{figure}

\section{Frame dragging errors during eclipse ``season''} As noted in Section 3, unbalanced out-of-plane forces can arise as the satellite enters eclipse ``season'', and in the opposite direction as it exits eclipse season. These generate torques which can result in an instantaneous nonzero $d\Omega/dt$ even if the orbital ellipticity is zero. This will become clearer as we describe the satellite orbit.

\noindent To describe the orbit, we will work in the celestial coordinate system\footnote{i.e. Cartesian coordinate system, with the Earth's equatorial plane as the x-y plane, the z-axis the axis of the Earth pointing roughly toward Polaris,
and the x-axis pointing toward the vernal equinox.}. The unit vector $\hat{r}_{Sun}$ pointing from Earth to Sun $k$ days after the vernal equinox is:
\begin{dmath}
\hat{r}_{Sun}{(k)} = \begin{pmatrix}
\cos{(2\pi k/A)}
\cos{\beta}\sin{(2\pi k/A)}
\sin{\beta}\sin{(2\pi k/A)}
\end{pmatrix}
\end{dmath}
where $A\approx 365.25$ is the length of the year in days, and $\beta = 23.2^\circ$ is the obliquity of the Sun's orbital plane (i.e. the angle this plane makes with the Earth equatorial plane). For simplicity we assume that for any given day, the position of the Sun in the sky and $\Omega$ stay fixed (i.e. they increment discretely from one day to the next\footnote{We shall see that this approximation suffices for our purposes.}). Also, the unit vector $\hat{r}_{sat}$ from the center of Earth to the center of the satellite is:
\begin{dmath}
\hat{r}_{sat}{(t,k)} = \begin{pmatrix}
-\sin{\Omega}\cos{i}\sin{\omega_{o} t}+\cos{\Omega}\cos{\omega_{o} t}\cos{\Omega}\cos{i}\sin{\omega_{o} t}+\sin{\Omega}\cos{\omega_{o} t}\sin{i}\sin{\omega_{o} t}
\end{pmatrix}
\end{dmath}
where $\omega_{0}$ is the satellite's orbital frequency, and $\Omega{(k)}$ is the longitude of the ascending node on day $k$. The time $t$ will be taken to range in $t \in [0,T]$, i.e. one orbital period, with the satellite at the longitude of the ascending node at $t=0$.\\

Finally, we need the orbital angular momentum. The period of LAGEOS (and of LARES 2) is $\approx 226 $ minutes, and its orbital radius (and LARES 2's) is $ 12,271 km$. The angular momentum magnitude per unit mass for both LARES 2 and LAGEOS is $ 12,271 \times 12,271 \times 2\pi/(226 \times 60 )\times 10^6 m^2/sec\approx 70\times 10^{9} m^2/sec.$ The direction of $L$ is given by:

\begin{dmath}
\hat{L}= \begin{pmatrix}
\sin{\Omega} \sin i
-\cos{\Omega}\sin i
\cos i
\end{pmatrix}.
\end{dmath}
This results from starting with an orientation with $\hat L$ along the $\hat z$ direction, then rotating $\hat L$ counterclockwise about $\hat x$ by the inclination angle $i$, then counterclockwise about $\hat z$ by the value of the argument of the ascending node, $\Omega$.

An instantaneous acceleration is generated on the satellite by the solar radiation force (the radiation pressure multiplied by the cross sectional area of the satellite) as discussed above.  The radiation force per unit mass $a_{s} \approx 3.75 \times^{-9} N/kg$ on LAGEOS. LARES 2 will have a cross sectional area $\approx 0.45$ that of LAGEOS, and we assume similar surface behavior for LARES 2 as for LAGEOS.  LAGEOS has mass 406kg, while  LARES 2's mass is expected to be 300kg, so, $a_{S} \approx 2.0 \times 10^{-9} N/kg$ on LARES 2.  The direction of the force is parallel but opposite $\hat r_{Sun}$. Hence the instantaneous torque is proportional to the cross product $-\hat{r}_{sat}{(t,k)} \times \hat{r}_{Sun}{(k)}$. That is, in components the unit vector $(\hat N_x, \hat N_y, \hat N_z)$ in the direction of the torque is:

\begin{dmath}
-\hat N_x = \cos{\beta}\sin{(2\pi k/A)} \sin{i}\sin{\omega_{o} t} - (\cos{\Omega}\cos{i}\sin{\omega_{o} t}+\sin{\Omega}\cos{\omega_{o} t})\sin{\beta}\sin{(2\pi k/A)}
\end{dmath}

\begin{dmath}
-\hat N_y = \sin{\beta}\sin{(2\pi k/A)}(-\sin{\Omega}\cos{i}\sin{\omega_{o} t}+\cos{\Omega}\cos{\omega_{o} t}) - \cos{(2\pi k/A)}\sin{i}\sin{\omega_{o} t}
\end{dmath}

\begin{dmath}
-\hat N_z = \cos{(2\pi k/A)}(\cos{\Omega}\cos{i}\sin{\omega_{o} t}+\sin{\Omega}\cos{\omega_{o} t}) - ( -\sin{\Omega}\cos{i}\sin{\omega_{o} t}+\cos{\Omega}\cos{\omega_{o} t})\cos{\beta}\sin{(2\pi k/A)}.
\end{dmath}

\noindent If there are no eclipses, this instantaneous torque averages to zero over the circular orbit because of cancellation from forces applied on opposite sides of the orbit. To consider eclipses we will need to compute the time of eclipse entrance $t_{entrance}$ and eclipse exit $t_{exit}$. To do this, we compute the perpendicular distance $d{(t,k)}$ from the satellite to the Earth-Sun axis (this is another application of the just computed torque direction):
\begin{dmath}
d{(t,k)} = r_{sat}|\hat{r}_{sat}{(t,k)} \times \hat{r}_{Sun}{(k)}|.
\end{dmath}
Then $t_{entrance}$ and $t_{exit}$ are solutions to the equation :
\begin{dmath}
d{(t,k)} = R_{E},
\end{dmath}
where $R_{E}$ is the Earth's radius. However, in general there will be 4 solutions. To see this, one can visualize a cylinder with the Earth-Sun axis for axis of symmetry, and with radius equal to the radius of Earth. The part of the cylinder ``behind'' the Earth is in shadow. Thus if, on a particular day, LARES experiences an eclipse, then its orbit will intersect the cylinder in 4 points, 2 of which are $t_{entrance}$ and $t_{exit}$. To pick out the correct two solutions, we consider the dot product $\hat{r}_{Sun} \cdot \hat{r}_{sat}$. When this dot product is negative, the vector $\hat{r}_{Sun}$ makes an angle larger than $\frac{\pi}{2}$ with the vector $\hat{r}_{sat}$, and there is eclipse.\\

 Since the orbit-average torque is orthogonal to $\hat L$, the change in length $|L|$ vanishes and we have
\begin{dmath}
\frac{dL}{dt} = m r_{sat}^2 \omega_o \frac{d \hat L}{dt},
\end{dmath}
which leads to the cross product
\begin{dmath}
\frac{d \hat L}{dt} = \frac{\hat r_{sat}}{\omega_0 |r_{sat}|} \times (F/m) =
 \frac{|F|}{mv_0}\hat r_{sat} \times (-\hat r_{Sun}) .
\end{dmath}

\noindent Here $v_0$ is the orbital velocity, $5690m/sec$ for LAGEOS and LARES 2. The factor $|F/m|/v_0$ evaluates to $\approx 6.5 \times 10^{-13} sec^{-1}$ for LAGEOS and about half that for LARES. We coded these equations for the LAGEOS and LARES 2 satellites and computed the effect on the nodal precession during the eclipse season.\\

Equation (12) was integrated using Mathematica's analytic integration capability. The results are shown in Figures 6-9. To understand these results, we reiterate the following points: Eq.(12) is the small motion of the direction of the angular momentum specifically due to the eclipses, ``on top of'' the much larger Newtonian and frame-dragging effects. Our target value for frame dragging error is $0.2\%$, of $\approx 30mas/yr$, i.e.  $0.06 mas/yr = 3 \times 10^{-7} radians/yr$. The computed components of $d \hat L/dt$ are angular rates ($radian/sec$).

The calculations demonstrated in Figures 6-9 show for both satellites first the components of $d \hat L/dt$, and then the time integrated value of $ \Delta \hat L$ as a function of time up to 10 years. The (artificially chosen) initial conditions have the line of nodes at $\Omega= 0$, i.e. aligned with the vernal equinox at the same time the sun is at the vernal equinox (the $V$-axis, which we also call the $x$-axis).
While this seems special, the motion of the sun and the satellite quickly move away from this special orientation.\footnote{This initial orientation means that the system starts at the middle of an eclipse season, and only half of the season is included in the displayed evolution. This has the effect of producing an initial step in the accumulated $\Delta \hat L$, which, however will be seen to be irrelevant in the discussion.}\\

Figures 6 and 7 are for the satellite LAGEOS. They are the instantaneous values of components of $d \hat L/dt$ along the north ($N$- also called $z$- axis), vernal ($V$- also called $x$- axis), and the third  ($N \times V$ also called $y$- axis), followed by the time-integrated value of $ \Delta \hat L$ for each of the same components, in the same order. We clearly see the separated episodes of change of direction during each eclipse season.

Figures 8 and 9 are the equivalent plots for the satellite LARES 2. LARES 2 is proposed to have an inclination supplementary to LAGEOS', i.e. $i_{LARES 2} \approx 70^\circ .16$.
With that inclination the classical Newtonian precession of LARES 2 is {\itshape opposite} LAGEOS'. LARES 2 precesses $0^ \circ .345$ {\itshape westward} per day. Its eclipse seasons thus occur approximately every 135 days, much more frequently than LAGEOS', as can be seen in the graphs. Also, because of the smaller area/mass ratio of LARES 2, the solar acceleration, hence the torque per angular momentum, and ultimately the values of $d \hat L/dt$ and $\Delta \hat L$, have approximately half the size for LARES 2 that they have for LAGEOS. (Note that the scales on the LARES 2 graph are different from those on the LAGEOS graphs.)

Thus, we concentrate on Figures 6 an 7 for the satellite LAGEOS. Fig. 7 presents the accumulated $\Delta \hat L$. The ($N$) component shows values between zero and $- 10^{-7}$ over the ten year period (and would be between $\pm 0.5 \times 10^{-8}$ if the initial half season offset is removed. Thus over ten years, the average rate is less than $10^{-8} radian/yr$. The ($V$) direction for LAGEOS shows a secular drift of about $2 \times 10^{-8} radian/yr$ over the ten years. $\Delta \hat L$ for the  ($N \times V$) direction shows an initial jump to about $3 \times 10^{-8}$ and is about $5 \times 10^{-8}$. Thus it averages to about $5 \times 10^{-9}$ over ten years.

The similar accumulated $\Delta \hat L$ for LARES 2 show even smaller long-term averages.
In particular, the LARES 2 ($V$) component shows a secular drift which averages to about $10^{-8}$ over ten years. The other two components are very small, and have {\itshape values} after ten years of less than $10^{-8}$ $radian$ implying rates less than $2 \times 10  ^{-9} radian/yr$.

For both LAGEOS and LARES 2 the predicted orbital precessions due to thermal thrust eclipse-induced torques are at least an order of magnitude below the  $3 \times 10^{-7} radian/yr$ error budget.

\begin{figure}
\centering
\includegraphics[width=0.640\textwidth]{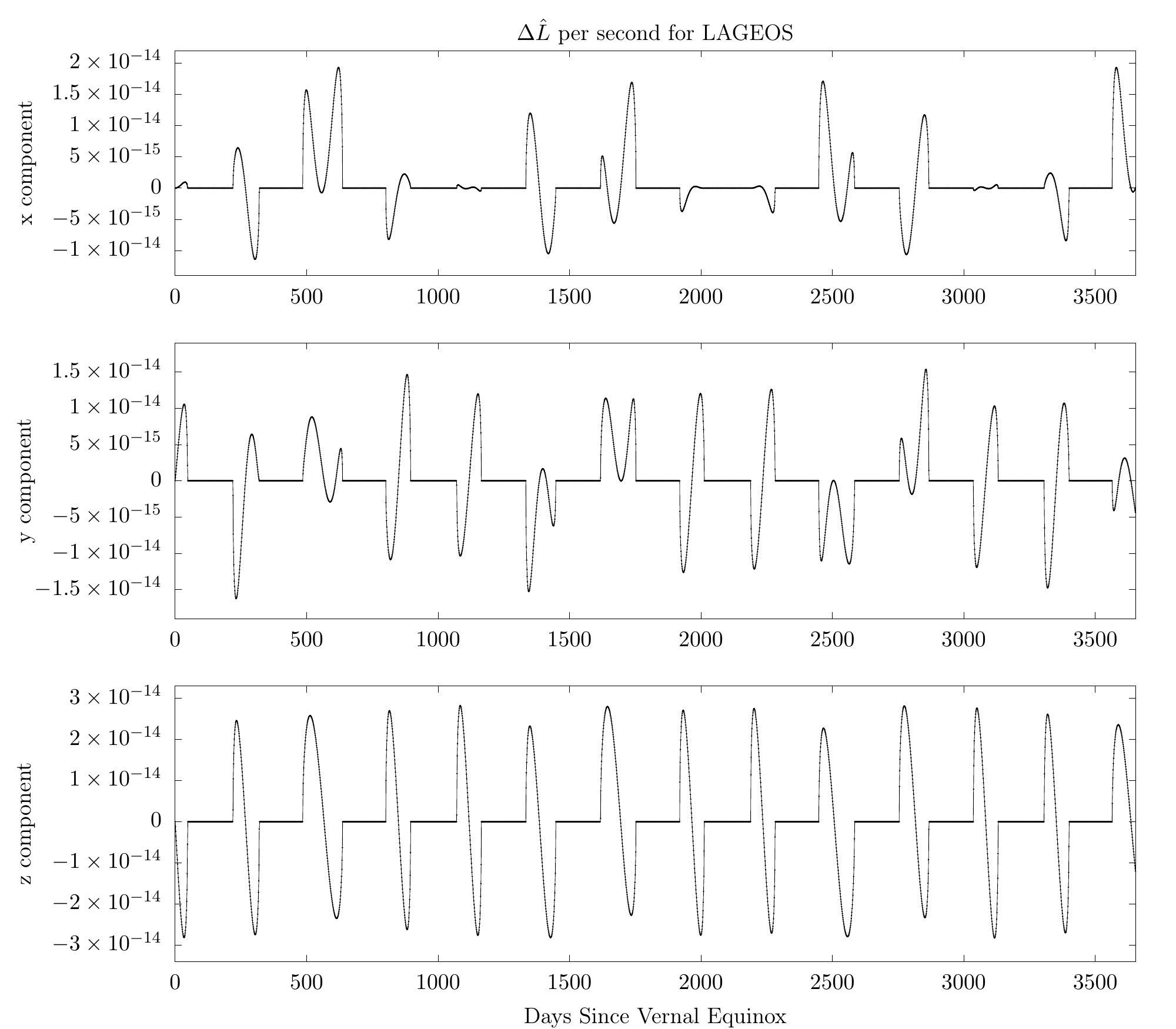}
\caption{Modeled instantaneous rate of change of orbital angular momentum direction $d\hat L/dt$ ($radians/sec$) due to eclipses for LAGEOS displayed over a 10 year period. The graphs refer to: $x-$component $\equiv$ direction toward $V$, the Vernal Equinox; $y-$ component in the $z\times x$ direction, towards $N \times V$; with the $z-$direction northward, ($N$). Note that the vertical scales are unique to each graph.}
\label{fig:6}       
\end{figure}

\begin{figure}
\centering
 \includegraphics[width=0.640\textwidth]{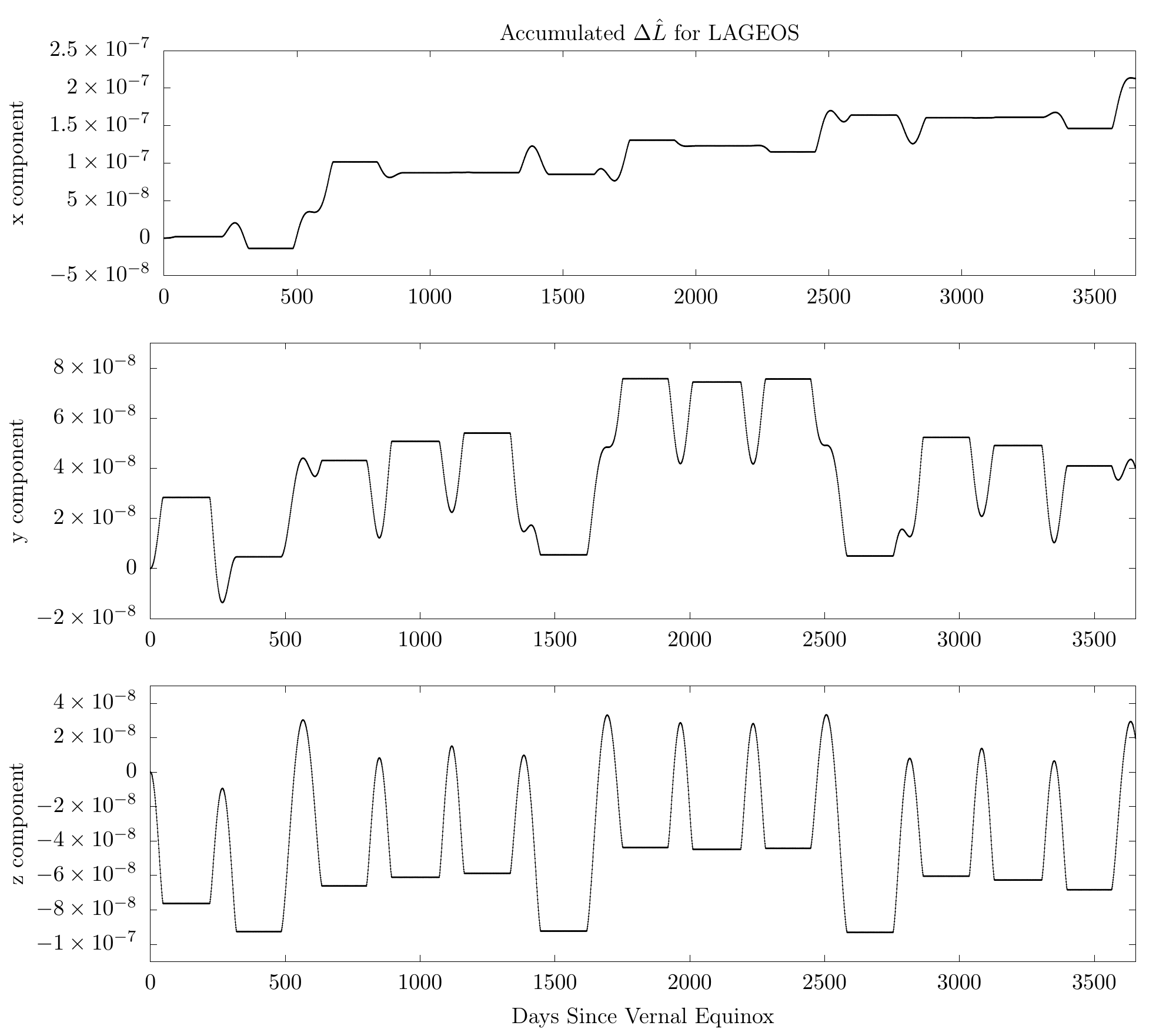}
\caption{Modeled accumulated value of eclipse contribution to LAGEOS orbital angular momentum direction change ($radians$) displayed over a 10 year period (time integrals of Fig 6 graphs). The graphs refer to: $x-$component $\equiv$ direction toward $V$, the Vernal Equinox; $y-$ component in the $z\times x$ direction, towards $N \times V$; with the $z-$direction northward, ($N$). Note that the vertical scales are unique to each graph.}
\label{fig:7}       
\end{figure}

\begin{figure}
\centering
 \includegraphics[width=0.640\textwidth]{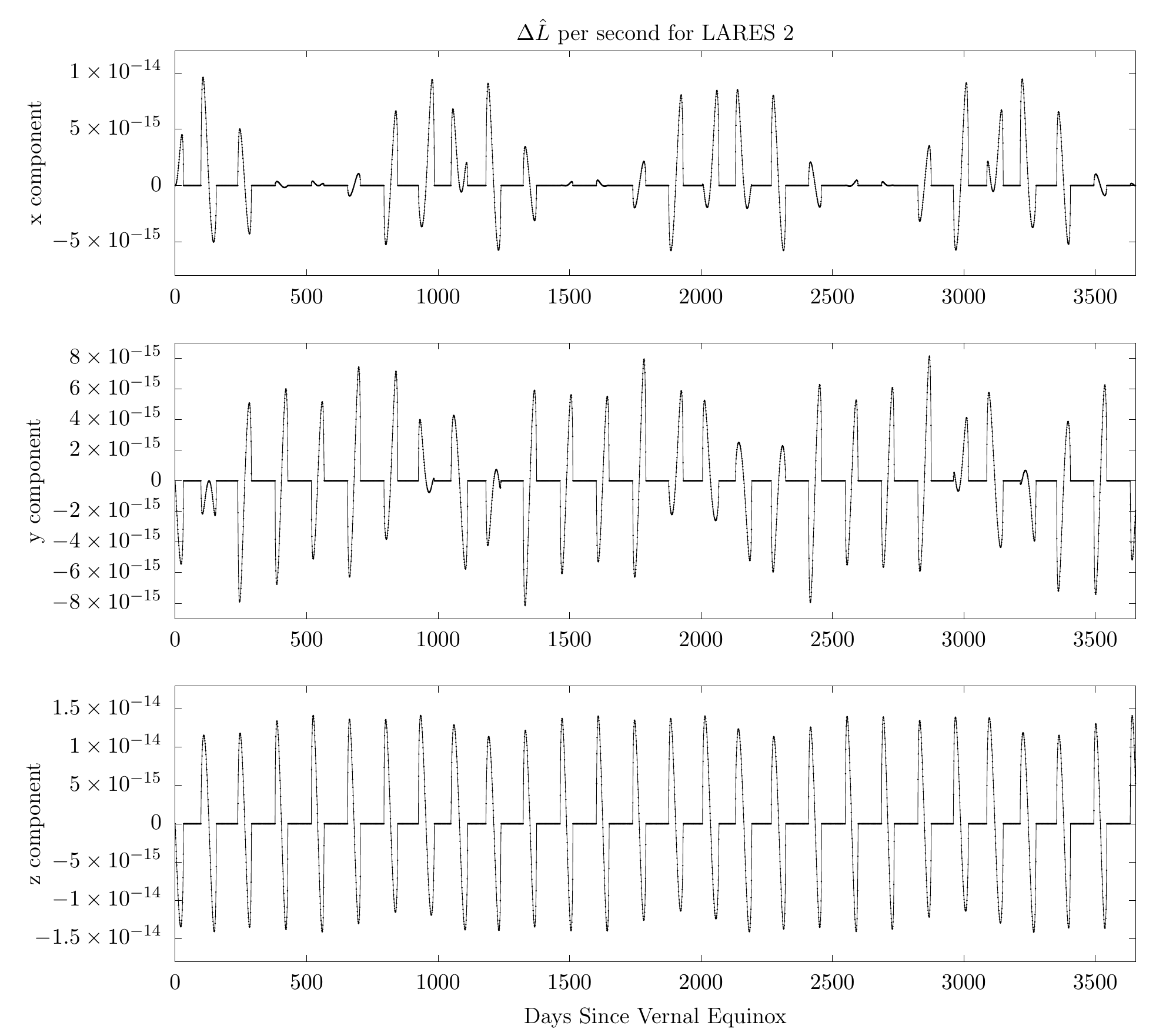}
\caption{Modeled instantaneous rate of change of orbital angular momentum direction $d\hat L/dt$ ($radians/sec$) due to eclipses for LARES 2 displayed over a 10 year period. The graphs refer to: $x-$component $\equiv$ direction toward $V$, the Vernal Equinox; $y-$ component in the $z\times x$ direction, towards $N \times V$; with the $z-$direction northward, ($N$). Note that the vertical scales are unique to each graph.}
\label{fig:8}       
\end{figure}

\begin{figure}
\centering
 \includegraphics[width=0.640\textwidth]{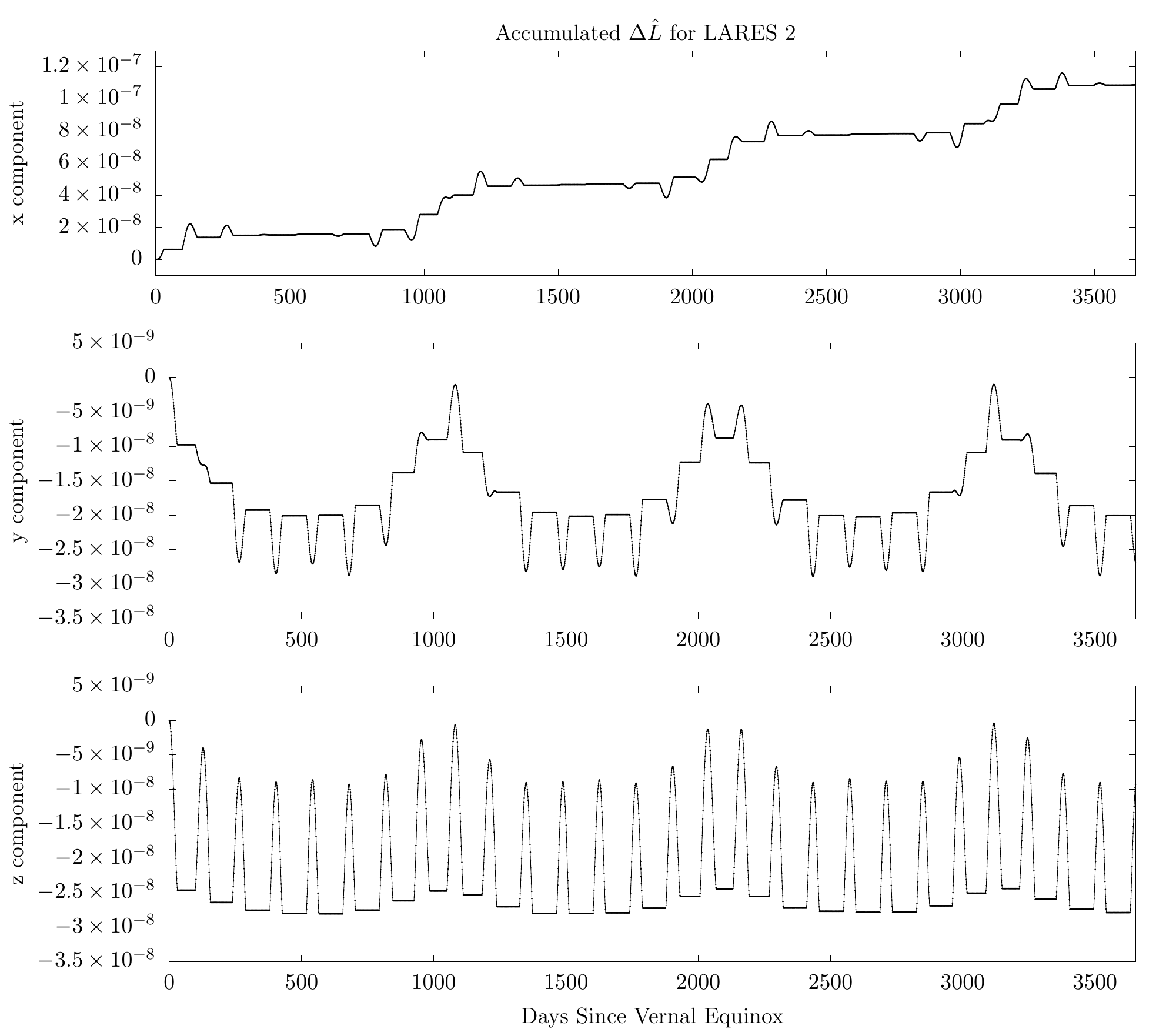}
\caption{Modeled accumulated value of eclipse contribution to LARES 2 orbital direction change ($radians$) displayed over a 10 year period (time integrals of Fig 8 graphs). The graphs refer to: $x-$component $\equiv$ direction toward $V$, the Vernal Equinox; $y-$ component in the $z\times x$ direction, towards $N \times V$; with the $z-$direction northward, ($N$). Note that the vertical scales are unique to each graph.}
\label{fig:9}       
\end{figure}

\section{Conclusions}

The final fractional error in the test of frame-dragging due to the Yarkovsky effect (solar radiation pressure torquing the orbit because the orbits have a slight eccentricity) on the satellites LAGEOS and LARES 2, considered to be rotationally at rest with respect to inertial space, will be between $0.3 \times 10^{-3}$ (simulations of thermal drag) and $1.47 \times 10^{-3}$ (data analysis of real data with different values of the $C_{r}$). These results agree with our estimated error \cite{bib9} of about  10$^{-3}$ in the test of frame-dragging with LAGEOS and LARES 2 due to the Yarkovsky effect in the case of non-spinning satellites. The predicted orbital precessions due to thermal thrust eclipse-induced torques are even smaller: at least an order of magnitude below the  $3 \times 10^{-7} radian/yr$ error budget, i.e. at the level of fractional errors of $10^{-4}$. These results are consistent with the anticipated frame dragging overall fractional accuracy of $2 \times 10^{-3}$.

\end{document}